\documentclass[12pt]{article}
\usepackage{epsf}
\usepackage{amsmath, amssymb}

\begin{document}

\newcommand{\rem}[1]{{\bf #1}}

\renewcommand{\thefootnote}{\fnsymbol{footnote}}
\setcounter{footnote}{0}
\begin{titlepage}

\def\thefootnote{\fnsymbol{footnote}}

\begin{center}

\hfill UT-11-44\\
\hfill IPMU-11-0206\\
\hfill December, 2011\\

\vskip .75in

{\Large \bf 

  Extra Matters Decree the Relatively Heavy Higgs of Mass 
  about 125 GeV in the Supersymmetric Model

}

\vskip .75in

{\large
Takeo Moroi, Ryosuke Sato and Tsutomu T. Yanagida
}

\vskip 0.25in

{\em Department of Physics, University of Tokyo,
Tokyo 113-0033, Japan
\\
and
\\
Institute for the Physics and Mathematics of the Universe,\\
University of Tokyo, Kashiwa 277-8568, Japan
}

\end{center}
\vskip .5in

\begin{abstract}

  We show that the Higgs mass about $125\ {\rm GeV}$ is easily
  realized in supersymmetric model with extra matters, simultaneously
  explaining the anomaly in the muon anomalous magnetic moment and the
  dark matter density.

\end{abstract}

\end{titlepage}

\renewcommand{\thepage}{\arabic{page}}
\setcounter{page}{1}
\renewcommand{\thefootnote}{\#\arabic{footnote}}
\setcounter{footnote}{0}

\section{Introduction}

We have not found any convincing and fundamental theory for explaining
the observed masses of quarks and leptons. However, it was pointed
out, long ago, that if one introduces a pair of extra matters, ${\bf
  10}$ and ${\bf \bar{10}}$, in supersymmetric (SUSY) standard model,
the Yukawa coupling $y_t$ for the top quark has a quasi infrared fixed
point at $y_t\sim 1$ \cite{Bardeen:1993rv}. This is an extremely
interesting observation, since we can understand the observed mass of
the top quark as a low-energy prediction of the theory. Furthermore,
it has been, recently, pointed out \cite{Asano:2011zt} that the extra
matters, ${\bf 10}$ and ${\bf \bar{10}}$, cancels anomalies of a
discrete $R$ symmetry in the SUSY standard model, provided that the
sum of their $R$ charges is zero.  (See also \cite{Kurosawa:2001iq}.)
This is itself very interesting in the LHC phenomenology, since they
naturally have a SUSY-invariant mass at the electro-weak scale through
the Giudice-Masiero mechanism \cite{Giudice:1988yz}. The further
crucial and important point is that a new Yukawa coupling $y_U$ for
the extra matters has also a quasi infrared fixed point at $y_U\simeq
1$ \cite{Martin:2009bg} which gives additional large contribution to
the mass of the lightest Higgs boson \cite{Moroi:1991mg,
  Moroi:1992zk}, and as a consequence the lightest Higgs boson
acquires naturally a relatively large mass of $120-130$ GeV
\cite{Martin:2009bg, Asano:2011zt, Endo:2011mc}.

Recently, the ATLAS and CMS experiments have announced the latest
results of their searches for Higgs boson.  In particular, the ATLAS
has found $3.6\sigma$ local excess of the Higgs-like signal near
$m_h\simeq 126\ {\rm GeV}$ \cite{ATLAS-CONF-2011-163}.  The CMS has
also observed more than $2\sigma$ local excess of the signal at
$m_h\simeq 124\ {\rm GeV}$ \cite{CMS-PAS-HIG-11-032}.  Although the
significances decrease once we take into account the so-called
``look-elsewhere effect,'' these excesses may be the indication of the
existence of the Higgs boson at around $m_h\simeq 125\ {\rm GeV}$.
(If we consider the global probability, the excesses are at
$2.3\sigma$ by ATLAS and at $1.9\sigma$ by CMS.)  If the Higgs mass is
as large as $\sim 125\ {\rm GeV}$ in the framework of minimal SUSY
standard model (MSSM), the mass scale of the superparticles are
required to be relatively high ($\gtrsim$ a few TeV) to enhance the
Higgs mass via radiative corrections \cite{Giudice:2011cg}.  However,
as we have mentioned, this is not the case if there exist extra
matters.

In this letter, taking the excesses of Higgs-like signals at around
$m_h\simeq 125\ {\rm GeV}$ seriously, we consider SUSY standard model
with extra matters.  We show that the Higgs mass of $m_h\simeq 125$
GeV is easily explained in the model with extra matters if their
masses are in a rage of $500$ GeV$-1$ TeV. Surprising is that there is
a wide range of parameter space in which not only the Higgs boson mass
but also the dark matter density and the anomalous magnetic dipole
moment (MDM) of the muon are simultaneously explained. We also show
that the gluino is most likely lighter than 1 TeV which can be tested
soon at the LHC.

\section{Model}

As we have mentioned in the introduction, we consider the SSM with
extra matters which are ${\bf 10}$ and ${\bf \bar{10}}$
representations in $SU(5)_{\rm GUT}$, which we denote ${\bf 10'}$ and
${\bf \bar{10}'}$, respectively.\footnote
{For another possibility, we can consider the model with three pairs
  of ${\bf 5}$ and ${\bf \bar{5}}$; for the enhancement of the Higgs
  mass, the same number of singlets should be also introduced.  In such a
  case, the quasi fixed point value of the Yukawa coupling constant
  relevant for the enhancement of the Higgs mass is smaller compared
  to the model with ${\bf 10}$ and ${\bf \bar{10}}$ representations.
  Even so, a significant enhancement of the Higgs mass may be possible
  in particular when the mass of the extra matters are relatively low.
  Detailed discussion on this case will be given elsewhere
  \cite{MorSatYan}.}

To discuss the low-energy phenomenology, we decompose the extra
matters as ${\bf 10'}=Q+U+E$ and ${\bf
  \bar{10}'}=\bar{Q}+\bar{U}+\bar{E}$, where $Q({\bf 3},{\bf 2},1/6)$,
$U({\bf \bar{3}},{\bf 1},-2/3)$, $E({\bf 1},{\bf 1},1)$, $\bar{Q}({\bf
  \bar{3}},{\bf 2},-1/6)$, $\bar{U}({\bf 3},{\bf 1},2/3)$, and
$\bar{E}({\bf 1},{\bf 1},-1)$ are gauge eigenstates of the
standard-model gauge group.  (The gauge quantum numbers for $SU(3)_C$,
$SU(2)_L$ and $U(1)_Y$ are shown in the parenthesis.)  Then, the
superpotential relevant for the discussion of the low-energy
phenomenology is
\begin{eqnarray}
  W = W^{\rm (MSSM)} + 
  y_U U Q H_u
  + M_U \bar{U} U
  + M_Q \bar{Q} Q
  + M_E \bar{E} E,
  \label{W}
\end{eqnarray}
where $W^{\rm (MSSM)}$ is the superpotential of the minimal SUSY
standard model (MSSM), $H_u$ is the up-type Higgs multiplet, $y_U$ is
the Yukawa coupling constant for the extra matters, while $M_U$,
$M_Q$, and $M_E$ are mass parameters whose origin is assumed to be the
Giudice-Masiero mechanism.  In our study, we neglect the
superpotential of the form ${\bf \bar{5}}_H\cdot{\bf
  \bar{10}'}\cdot{\bf \bar{10}'}$ (where ${\bf \bar{5}}_H$ denotes the
Higgs multiplet in ${\bf \bar{5}}$ representation contains down-type
Higgs) because it is not important for the following discussion.  In
addition, the soft SUSY breaking terms are given by\footnote
{We neglect the bi-linear SUSY breaking terms for extra matters for
  simplicity.}
\begin{eqnarray}
  {\cal L}_{\rm soft} &=& 
  m_{\tilde{Q}}^2 |\tilde{Q}|^2
  + m_{\tilde{\bar{Q}}}^2 |\tilde{\bar{Q}}|^2
  + m_{\tilde{U}}^2 |\tilde{U}|^2
  + m_{\tilde{\bar{U}}}^2 |\tilde{\bar{U}}|^2
  + m_{\tilde{E}}^2 |\tilde{E}|^2
  + m_{\tilde{\bar{E}}}^2 |\tilde{\bar{E}}|^2
  \nonumber \\ &&
  + (y_U A_U \tilde{U} \tilde{Q} H_u
  + {\rm h.c.}),
  \label{Lsoft}
\end{eqnarray}
where ``tilde'' is used for superpartners.  

The low energy parameters (in particular, the soft SUSY breaking
parameters) given above are related to fundamental parameters given at
high scale.  As in the case of conventional study, we assume that the
boundary condition of the low-scale parameters are given at the
so-called GUT scale $M_{\rm GUT}$, which we take $M_{\rm GUT}=2\times
10^{16}\ {\rm GeV}$.  Because the GUT is one of the strong motivation
to consider SUSY, we consider a boundary condition which respects
$SU(5)$ symmetry.  In order to reduce the number of free parameters
(as well as to avoid dangerous flavor problems), we adopt the
following boundary condition at the GUT scale:
\begin{itemize}
\item All the gaugino masses are unified to $m_{1/2}$ at the GUT scale.
\item All the matter fields in ${\bf \bar{5}}$ representation except
  Higgs have universal SUSY breaking mass-squared parameter, denoted
  as $m^2_{\bf \bar{5}}$.
\item All the matter fields in ${\bf 10}$ and ${\bf \bar{10}}$
  representations have universal SUSY breaking mass-squared parameter,
  denoted as $m^2_{\bf 10}$.
\item The up- and down-type Higgses have SUSY breaking mass-squared
  parameters $m^2_{{\bf 5}_H}$ and $m^2_{{\bf \bar{5}}_H}$,
  respectively.
\item For simplicity, we assume that the SUSY breaking tri-linear
  scalar couplings vanish at the GUT scale.  Notice that the following
  results are almost unchanged even if we relax this assumption, since
  the tri-linear scalar couplings become small at quasi infrared fixed
  points in the present model \cite{Martin:2009bg}.
\end{itemize}
Notice that, in the following, the above mass-squared parameters are
allowed to be negative, which is important to have a viable low-energy
phenomenology as we will see below.

Once the boundary condition is fixed, we solve the renormalization
group equation from $M_{\rm GUT}$ to $M_{\rm SUSY}$, where $M_{\rm
  SUSY}$ is the typical mass scale of the superparticles.  We mostly
use the one-loop $\beta$-functions.  However, the two-loop effects on
the gaugino masses (in particular, gluino mass) are significant, so we
also take into account $O(\alpha_3^2)$ (as well as $O(\alpha_1^2)$ and
$O(\alpha_2^2)$) contributions to the $\beta$-functions of gaugino
masses and gauge coupling constants.  Then, using the soft SUSY
breaking parameters at low energy scale, we solve the conditions of
electro-weak symmetry breaking to determine the SUSY invariant Higgs
mass (i.e., so-called $\mu$-parameter) as well as bi-linear SUSY
breaking Higgs mass parameter (i.e., so-called $B$-parameter); in our
analysis, we use the tree-level minimization condition of the Higgs
potential.

With the low-energy parameters given above, we calculate experimental
observables, in particular, the muon MDM $a_\mu$ and the relic density
of the lightest superparticle as well as the the lightest Higgs mass
$m_h$.  The relevant formula to calculate $a_\mu$ can be found in
\cite{Moroi:1995yh}.  For the calculation of the density parameter of
the LSP $\Omega_{\rm LSP}$, we use {\tt DarkSUSY} package
\cite{Gondolo:2004sc}.

For the calculation of the Higgs mass, we use the fact that the model
is well described by the standard model once the extra matters and
superparticles decouple.  Then, we denote the potential of the
standard-model like Higgs $H_{\rm SM}$ as
\begin{eqnarray}
  V_{\rm SM} = m_H^2 |H_{\rm SM}|^2 + \frac{1}{2} \lambda |H_{\rm SM}|^4,
\end{eqnarray}
and the Higgs mass is given by
\begin{eqnarray}
  m_h^2 = \lambda (m_h) v^2,
\end{eqnarray}
where $v\simeq 246\ {\rm GeV}$ is the vacuum expectation value of the
standard-model-like Higgs boson.  The quartic Higgs coupling $\lambda$
is determined by the parameters in the supersymmetric model at the
mass scale of superpartners and extra matters as
\begin{eqnarray}
  \lambda(M_{\rm SUSY}) = \frac{1}{4} (g_2^2 + g_1^2) \cos^2 2\beta
  + \delta \lambda' ,
  \label{lambda_MSUSY}
\end{eqnarray}
where $g_2$ and $g_1$ are gauge coupling constants for $SU(2)_L$ and
$U(1)_Y$, respectively, and $\delta\lambda'$ is the contribution from
the extra matters.  Then, $\lambda(M_{\rm SUSY})$ and $\lambda (m_h)$
are related by using the standard-model renormalization group
equations.  In our analysis, we estimate $\delta\lambda'$ from
one-loop effective potential obtained by integrating out the extra
matters.  Denoting such an effective potential as $\Delta V^{\rm
  (extra)}$, we obtain
\begin{eqnarray}
  \delta \lambda' = \frac{1}{2}
  \frac{\partial^4\Delta V^{\rm (extra)}}{\partial H_u^2\partial {H_u^*}^2}
  \sin^4 \beta.
\end{eqnarray}
The one-loop contribution of the extra matters to the Higgs potential
is given by
\begin{eqnarray}
  \Delta V^{\rm (extra)} =
  \Delta V^{\rm (extra)}_{\rm B} + \Delta V^{\rm (extra)}_{\rm F},
\end{eqnarray}
where $\Delta V^{\rm (extra)}_{\rm B}$ and $\Delta V^{\rm
  (extra)}_{\rm F}$ are contributions of bosonic and fermionic loops,
respectively.  $\Delta V^{\rm (extra)}_{\rm B}$ is given by
\begin{eqnarray}
  \Delta V^{\rm (extra)}_{\rm B} = 
  \frac{3}{32\pi^2} {\rm Tr}
  \left[ ({\cal M}_{\rm B}^2 + \Delta {\cal M}_{\rm B}^2)^2 
    \left\{ 
      \ln  \left( \frac{{\cal M}_B^2 
          + \Delta {\cal M}_{\rm B}^2}{\mu^2} \right)
      - \frac{3}{2} \right\}
  \right],
\end{eqnarray}
where
\begin{eqnarray}
  {\cal M}_{\rm B}^2 = {\rm diag}
  (M_Q^2 + m_{\tilde{Q}}^2, M_Q^2 + m_{\tilde{\bar{Q}}}^2, 
  M_U^2 + m_{\tilde{U}}^2, M_U^2 + m_{\tilde{\bar{U}}}^2),
\end{eqnarray}
and
\begin{eqnarray}
  \Delta {\cal M}_{\rm B}^2 = \left( \begin{array}{cccc}
      y_U^2 |H_u|^2 & 0 & 0 & y_U M_U H_u^* \\
      0 & 0 & y_U M_Q H_u^* & 0 \\
      0 & y_U M_Q H_u & y_U^2 |H_u|^2 & 0 \\
      y_U M_U H_u & 0 & 0 & 0 \\
    \end{array} \right),
\end{eqnarray}
while
\begin{eqnarray}
  \Delta V^{\rm (extra)}_{\rm F} = 
  - \left. \Delta V^{\rm (extra)}_{\rm B} \right|_{m_{\tilde{Q}}^2
    =m_{\tilde{\bar{Q}}}^2=m_{\tilde{U}}^2=m_{\tilde{\bar{U}}}^2=0}.
\end{eqnarray}
Notice that we are interested in the case where the tri-linear coupling
constants tend to go to the fixed-point values, which are significantly
smaller than the soft SUSY breaking mass-squared parameters.  Thus, in
our analysis, we neglect the effects of the tri-linear coupling
constants in the estimation of the Higgs mass.

\section{Higgs Mass, Muon MDM, and $\Omega_{\rm LSP}$}

Now we numerically calculate the Higgs mass, the muon MDM, and the
relic density of the LSP.  Our main purpose is to show that there
exists a parameter region where the observed Higgs mass ($\sim 125\
{\rm GeV}$) as well as the muon MDM and the dark matter abundance are
simultaneously explained in the present framework.  Thus, we fix some
of the parameters rather than performing a systematic scan of the full
parameter space.  (More complete analysis will be given elsewhere
\cite{MorSatYan}.)
Here, we take $m^2_{{\bf 5}_H}=m^2_{{\bf \bar{5}}_H}=0$.
Because we would like to enhance the SUSY
contribution to the muon MDM, we adopt a large value of $\tan\beta$;
numerically, we take $\tan\beta=50$.
In addition, taking account of
the quasi fixed point behavior of the Yukawa coupling constant, we
take $y_U(M_{\rm SUSY})=1$.  (Here, we take $M_{\rm SUSY}=1.5\ {\rm
  TeV}$, which is the typical mass scale of the superparticle in the
following analysis.  The results given below are quite insensitive to
this value.)  Furthermore, we approximate $M_Q=M_U$ for simplicity.

\begin{figure}[t]
  \centerline{\epsfxsize=0.5\textheight\epsfbox{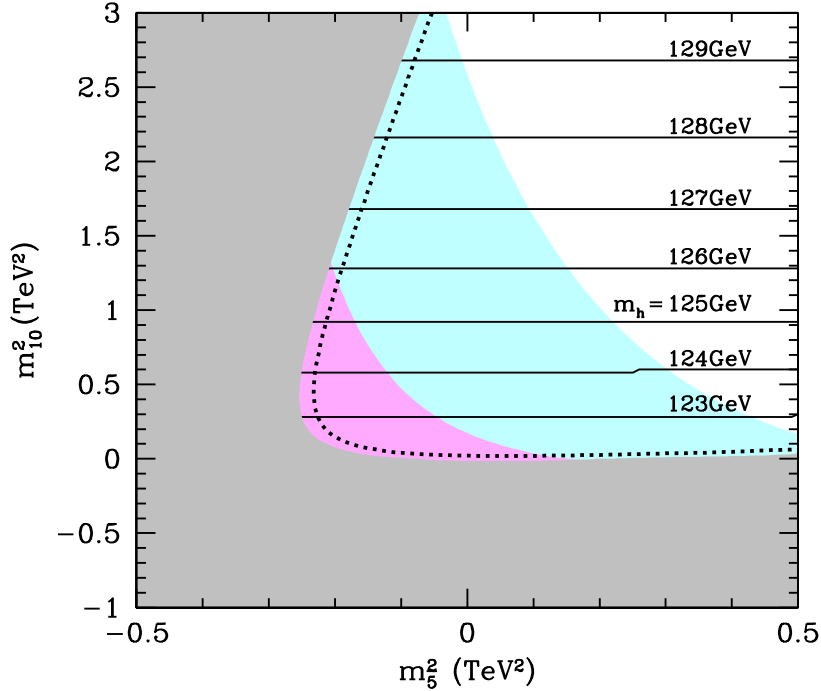}}
  \caption{\small Contours are those of constant $m_h$ ($123\ {\rm
      GeV}$, $\cdots$, $129\ {\rm GeV}$, from below) on $m_{\bf
      {\bar{5}}}^2$ vs.\ $m_{\bf 10}^2$ plane.  In the gray-shaded
    region, the lightest slepton mass becomes smaller than $100\ {\rm
      GeV}$.  On the dotted line, the lightest slepton mass becomes
    equal to the lightest neutralino mass.  In the pink (blue) region,
    the muon MDM becomes consistent with the experimental value at
    $1\sigma$ ($2\sigma$) level.  Here, we take $\tan\beta=50$,
    $M_U=550\ {\rm GeV}$, and $m_{1/2}\simeq 1.1\ {\rm TeV}$ (which
    corresponds to the Bino, Wino, and gluino masses of $170\ {\rm
      GeV}$, $240\ {\rm GeV}$, $700\ {\rm GeV}$, respectively).}
  \label{fig:exg700}
\end{figure}

\begin{figure}[t]
  \centerline{\epsfxsize=0.5\textheight\epsfbox{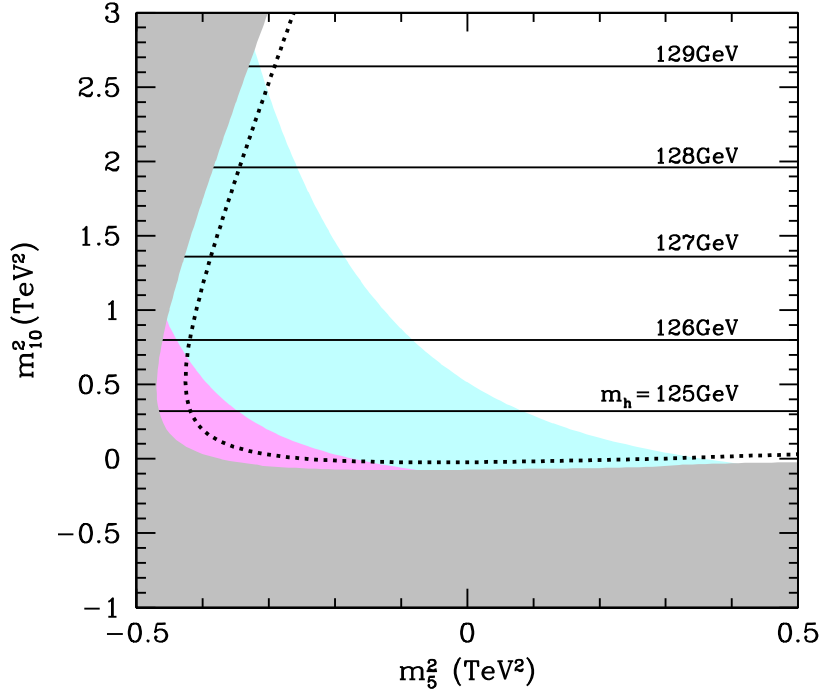}}
  \caption{\small Contours are those of constant $m_h$ ($125\ {\rm
      GeV}$, $\cdots$, $129\ {\rm GeV}$, from below) on $m_{\bf
      {\bar{5}}}^2$ vs.\ $m_{\bf 10}^2$ plane.  In the gray-shaded
    region, the lightest slepton mass becomes smaller than $100\ {\rm
      GeV}$.  On the dotted line, the lightest slepton mass becomes
    equal to the lightest neutralino mass.  In the pink (blue) region,
    the muon MDM becomes consistent with the experimental value at
    $1\sigma$ ($2\sigma$) level.  Here, we take $\tan\beta=50$,
    $M_U=650\ {\rm GeV}$, and $m_{1/2}\simeq 1.3\ {\rm TeV}$ (which
    corresponds to the Bino, Wino, and gluino masses of $210\ {\rm
      GeV}$, $310\ {\rm GeV}$, $900\ {\rm GeV}$, respectively).}
  \label{fig:exg900}
\end{figure}

In Figs.\ \ref{fig:exg700} and \ref{fig:exg900}, we show the results
of our numerical calculations, where we take $m_{1/2}\simeq 1.1$ and
$1.3\ {\rm TeV}$, respectively.  (The predicted Bino, Wino, and gluino
masses are $170\ {\rm GeV}$, $240\ {\rm GeV}$, $700\ {\rm GeV}$, and
$210\ {\rm GeV}$, $310\ {\rm GeV}$, $900\ {\rm GeV}$, respectively.)
In Fig.\ \ref{fig:exg700} (Fig.\ \ref{fig:exg900}), we take $M_U=550\
{\rm GeV}$ ($650\ {\rm GeV}$); these values of the mass of extra
generation quarks are above the present experimental lower bound on
the mass of fourth-generation quarks \cite{Luk:2011np}.  In the
gray-shaded region, the stau becomes lighter than $100\ {\rm GeV}$.
(In fact, in most of the gray-shaded region, the lighter stau becomes
tachyonic.)  In addition, on the dotted line, the stau and the
lightest neutralino become degenerate.  Thus, the stau is the LSP in
the region between the dotted line and the gray-shaded region while
the lightest neutralino (which is almost purely Bino) is the LSP in
the region right to the dotted line.  It is notable that the allowed
parameter region extends to the region with negative mass-squared
parameter at the GUT scale.  This is due to the fact that, in the
present model, the gaugino masses are more enhanced at high scale
compared to the case of the MSSM.  Consequently, gaugino contributions
to the renormalization group running of the scalar masses are
significantly enhanced.  For the left-handed sleptons, such an
enhancement is so significant that the masses of the left-handed
sleptons (at $M_{\rm SUSY}$) can be positive even if $m_{\bf
  \bar{5}}^2<0$.  Notice that, even if some of the scalar masses are
negative at the GUT scale, they become positive at the scale orders of
magnitude larger than $M_{\rm SUSY}$.  Thus, the tunneling to the
unwanted true vacuum is strongly suppressed and irrelevant.

In the figures, the (almost) horizontal lines are contours of constant
Higgs mass.  As we mentioned earlier, the Higgs mass can be as large
as $125\ {\rm GeV}$ in the present framework.  The Higgs mass is more
enhanced for larger value of $m_{\bf 10}^2$.  This is due to the fact
that the leading contribution to the Higgs mass from the extra matters
is approximately proportional to $\log(m_{\tilde{U}}/M_U)$.  This fact
also indicates that, if we increase (decrease) $M_U$, the Higgs mass
becomes smaller (larger).

In the same figures, we also show the region where the SUSY
contribution to the muon MDM $\Delta a_\mu^{\rm (SUSY)}$ well explains
the $\sim 3\sigma$ discrepancy between the experimental and
standard-model values of the muon MDM \cite{Cho:2011rk},
\begin{eqnarray}
  \Delta a_\mu^{\rm (SUSY)} =
  a_\mu^{\rm (exp)}  - a_\mu^{\rm (SM)} 
  = (25.9 \pm 8.1)\times 10^{-10},
  \label{amu(exp)}
\end{eqnarray}
where $a_\mu^{\rm (exp)}$ and $a_\mu^{\rm (SM)}$ are experimental
value and standard-model prediction of $a_\mu$.  One can see that, in
the model with extra matters, the lightest Higgs mass can be $\sim
125\ {\rm GeV}$ with satisfying the muon MDM constraint (at the
$1\sigma$ level).
 To understand the behavior of the SUSY
  contribution to the muon MDM, one should note that $\Delta
  a_\mu^{\rm (SUSY)}$ is from smuon-neutralino and sneutrino-chargino
  loop diagrams and that $\Delta a_\mu^{\rm (SUSY)}$ is approximately
  proportional to $\tan\beta$.  Thus, in order to make $\Delta
  a_\mu^{\rm (SUSY)}$ sizable, at least one of the left- or
  right-handed smuon should be relatively light.  As we have
  discussed, by taking $m_{\bf \bar 5}^2<0$, the left-handed slepton
  masses can be small even though the renormalization group effect on
  the sfermion masses is significant.
Then, with our present choice of $\tan\beta=50$ and the light slepton masses,
we obtain sufficient $\Delta a_\mu$ to realize Eq.\ \eqref{amu(exp)}.

Next, we discuss the thermal relic density of the LSP.  In particular,
on the right of dotted line, the lightest neutralino is the LSP, so it
is a viable candidate of dark matter.  However, in the bulk of such a
region, sleptons are much heavier than the lightest neutralino.  In
addition, the lightest neutralino is (almost) purely Bino.  Thus, the
pair annihilation cross section of the lightest neutralino is so small
that $\Omega_{\rm LSP}$ becomes too large to be consistent with the
present dark matter density if there is no other annihilation process.
As we have mentioned, in the parameter region near the dotted line,
the lightest neutralino almost degenerates with the stau and the
co-annihilation process becomes important \cite{Drees:1992am,
  Mizuta:1992qp}.  In particular, if the mass difference between the
lightest neutralino and the stau is a few GeV, the relic density of
the lightest neutralino becomes consistent with the present dark
matter density $\Omega_ch^2=0.1116$ (with $h$ being the Hubble
parameter in units of km/sec/Mpc) \cite{Komatsu:2010fb}.  We have
checked that there indeed exists a contour on which $\Omega_{\rm
  LSP}h^2=0.1116$ so that the lightest neutralino can be dark matter.
If we draw such a contour on the figure, it is (almost)
indistinguishable from the dotted line, and it crosses the contour of
$m_h=125\ {\rm GeV}$ in the region where the muon MDM anomaly can be
explained at $1\sigma$ level.  A similar situation was also studied in
\cite{Endo:2011mc}, where the simultaneous explanation of $m_h=125\
{\rm GeV}$, the muon MDM anomaly at $1\sigma$ level, and the dark
matter density was hardly realized in the so-called mSUGRA model.
Compared to \cite{Endo:2011mc}, our result crucially depends on the
fact that we allowed soft SUSY breaking mass-squared parameters to be
negative.

Finally, we comment on the gluino mass constraint.  In the limit of
heavy squark masses, the present experimental lower bound on the
gluino mass is $\sim 700\ {\rm GeV}$ \cite{ATLAS-CONF-2011-155}.
(Here, we have used the constraints on so-called
``squark-gluino-neutralino model.'')  On the first sample point used
in our numerical calculation, the predicted gluino mass is marginally
consistent with the bound.  In other points, like our second sample
point, $m_h\simeq 125\ {\rm GeV}$, the muon MDM, and the dark matter
density can be simultaneously explained with heavier gluino mass.
Even in such a case, the gluino mass can be well within the reach of
future LHC experiment.  We comment here that, with very large gluino
mass, it becomes difficult to solve the muon MDM anomaly.  This is
because, adopting larger value of $m_{1/2}$, the renormalization
effect on the SUSY breaking mass-squared parameter of the up-type
Higgs boson becomes more negative.  Then, in order to realize the
proper electro-weak symmetry breaking, a large value of the
$\mu$-parameter is needed, resulting in the suppression of the muon
MDM.  Thus, the search for the gluino signal at the LHC is a crucial
test of the present scenario.  We also note here that the muon MDM can
be enhanced if our assumptions on the mass spectrum of superparticles
are relaxed.  For example, with the gluino mass being fixed, the muon
MDM becomes larger if the Wino and Bino masses are somehow suppressed.
Such a mass spectrum is possible if the GUT relation among the gaugino
masses is violated; in the product group GUT scenario
\cite{Yanagida:1994vq}, this may be the case
\cite{ArkaniHamed:1996jq}.

\section{Summary}

In this letter, we have argued that the Higgs mass of $\sim 125\ {\rm
  GeV}$, around which ATLAS \cite{ATLAS-CONF-2011-163} and CMS
\cite{CMS-PAS-HIG-11-032} have observed excesses of Higgs-like
signals, can be well explained in SUSY models with extra matters.  In
the MSSM, it is often the case where large values of superparticle
masses are preferred to realize such a value of the Higgs mass.
However, such a mass spectrum tends to suppress the SUSY contribution
to the muon MDM, and also makes the LHC searches for the
superparticles difficult.  However, in the present model, $m_h\simeq
125\ {\rm GeV}$ is realized in the region where the SUSY contribution
to the muon MDM can be large enough to explain the $\sim 3\sigma$
discrepancy between the experimental and standard-model values of the
muon MDM (at $1\sigma$).  Simultaneously, the relic density of the
lightest neutralino can become consistent with the present dark matter
density.  We have seen that, for the the simultaneous explanation of
the Higgs mass, the muon MDM, and the dark matter density, some of the
SUSY breaking mass-squared parameters is preferred to be negative at
the GUT scale.  In the parameter region we are interested in, the
gluino mass as well as the masses of extra matters can be below $\sim
1\ {\rm TeV}$, so they are within the reach of future LHC experiments.

{\it Acknowledgments:} This work is supported by Grant-in-Aid for
scientific research from the Ministry of Education, Science, Sports,
and Culture (MEXT), Japan, No.\ 22244021 (T.M. and T.T.Y.), No.\
22540263 (T.M.), No.\ 23104001 (T.M.), and also by the World Premier
International Research Center Initiative (WPI Initiative), MEXT,
Japan. The work of R.S. is supported in part by JSPS Research
Fellowships for Young Scientists.

\end{document}